# Addressing systematic errors in axial distance measurements in single-emitter localization microscopy

PETAR N. PETROV AND W. E. MOERNER*

*Department of Chemistry, Stanford University, 333 Campus Drive, Stanford, CA 94305, USA*
*wmoerner@stanford.edu*

**Abstract:** Nanoscale localization of point emitters is critical to several methods in optical fluorescence microscopy, including single-molecule super-resolution imaging and tracking. While the precision of the localization procedure has been the topic of extensive study, localization accuracy has been less emphasized, in part due to the challenge of producing an experimental sample containing unperturbed point emitters at known three-dimensional positions in a relevant geometry. We report a new experimental system which reproduces a widely-adopted geometry in high-numerical aperture localization microscopy, in which molecules are situated in an aqueous medium above a glass coverslip imaged with an oil-immersion objective. We demonstrate a calibration procedure that enables measurement of the depth-dependent point spread function (PSF) for open aperture imaging as well as imaging with engineered PSFs with index mismatch. We reveal the complicated, depth-varying behavior of the focal plane position in this system and discuss the axial localization biases incurred by common approximations of this behavior. We compare our results to theoretical calculations.

## 1. Introduction

In recent years, single-emitter localization microscopy has emerged as a powerful platform for fluorescent bioimaging. By fitting the point spread function (PSF) of an emitter to a model function, one can obtain an estimate of the two-dimensional position of the emitter with sub-diffraction precision. Combined with control of the emitting concentration to avoid emitter overlap, this enables the formation of super-resolution images of static structures and the precise reconstruction of single-particle trajectories.

While localization microscopy is a popular tool for studying single biomolecules on a glass surface or on the membranes of adherent cells [1-5], it can also be extended to three-dimensional (3D) imaging by taking advantage of the axial ($z$) dependence of the PSF. Multiplane, interferometric, and PSF engineering methods have been used to perform 3D localization, with each approach conferring advantages and disadvantages [6]. All of these methods benefit from the use of high-numerical aperture (NA) objectives to collect as much light from the molecule as possible and reduce the diffraction-limited PSF size. Typically, oil immersion objectives such as the popular NA 1.4/100x magnification objective are used in order to maximize NA.

Importantly, bioimaging experiments are conducted largely in aqueous media with a refractive index of ~1.33, while the glass coverslip on which samples are mounted and the oil that lies beneath it (Fig. 1) have a refractive index of e.g. ~1.52. This refractive index mismatch within the sample geometry enables the collection of supercritical fluorescence, a bright contribution from near-field emission converted to propagating far-field light, from emitters close to the water-glass interface [7,8]. The significant increase in photon counts from emitters near the surface is an advantage of this experimental geometry, and supercritical emission can even be used to perform 3D localization [9-11]. However, the mismatch also creates a disturbance in the wavefront as fluorescence propagates across the water-glass interface and enters the objective lens. The wavefront distortion can be analytically modeled as a spherical aberration which intensifies with increasing depth of the emitter into the aqueous medium, away from the water-glass interface [12-14]. The effect of this spherical aberration on the PSF shape is, qualitatively, to reduce the peak intensity by spreading photons over a larger area and to distort the axial cross-section such that the PSF becomes increasingly asymmetric with respect to its focal position [15-17]. The loss of contrast that results from the aberration of the PSF can be addressed through the use of adaptive optics, which allow a compensatory spherical phase to be added to the optical system to counteract the PSF distortion [18-22].

Another side effect of the spherical aberration is that the focal position, at which the PSF is the most tightly focused and has the smallest waist, becomes displaced with respect to the axial position of the emitter. This has been termed the focal shift of the optical system, and describes the observation that the "nominal" focal position of the microscope (i.e., the relative position of the sample stage and objective lens; $\Delta f$ in Fig. 1) and the actual focal position (i.e., the true location within the sample of emitters that appear "in focus"; $z$ in Fig. 1) are displaced. The displacement between the nominal and actual focus grows as the focal plane moves further into the aqueous medium. Particularly in confocal and deconvolution microscopies, this effect has undergone considerable study at the scale of tens to hundreds of microns above the coverslip [16,23-27].

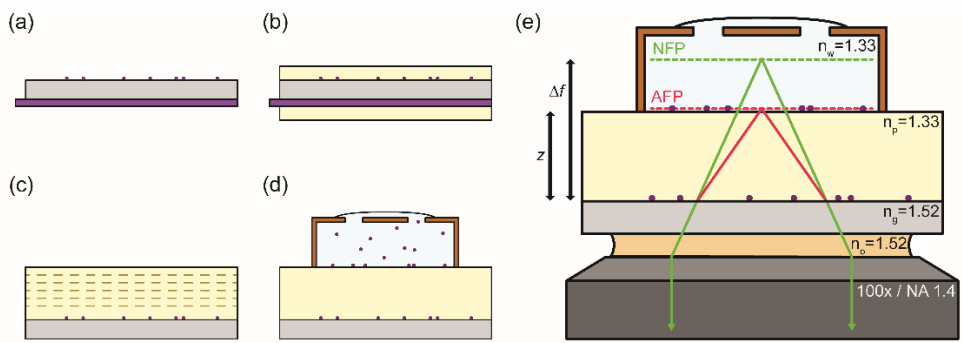

Fig. 1. Sample preparation protocol. (a) Beads are spin-coated onto the top side of the coverslip while a protective coating is applied to the bottom. (b) Dip-coating of the sample in polymer solution produces coatings on both sides of the coverslip. (c) Protective coating is removed, exposing the bottom surface of the coverslip. Dashed lines show additional polymer coats added to the top layer via spin-coating. (d) Chamber is added and filled with aqueous bead solution. (e) Chamber is washed with water to remove floating beads. The final configuration is shown. Refractive index mismatch causes rays emitted from the actual focal plane (AFP) to bend at the glass-polymer interface, thus appearing to have come from the nominal focal plane (NFP). Refractive indices of all media in the optical path are indicated.

The effects of PSF distortion and focal shift are particularly important to consider when performing localization microscopy in a refractive index mismatch situation. Depth-dependent distortion of the PSF by spherical aberration limits the quality of PSF models and the utility of calibration procedures performed at the coverslip surface (or any other single $z$ position) [28-33], while axial focal shift requires rescaling of the axial position estimates from nominal to true values [34,35]. These complications can hamper both the precision and accuracy of 3D localization if not accounted for properly. Unfortunately, creating an experimental geometry in which stationary, unperturbed, sub-diffraction fluorescent emitters are imaged at known axial positions within several microns of the water-glass interface has posed a challenge. Past efforts have relied upon tethering emitters to high-index structures which risk perturbation of the PSF [36,37], using large reference objects rather than single sub-diffraction emitters [38,39], using emitters dispersed randomly in samples lacking ground truth $z$ information [31,33,40,41], and using an optical trap reliant on optical readout of the ground truth $z$ [42,43].

Here, we demonstrate an experimentally useful model for studying the water-glass interface based on a polymer spacer matched to the refractive index of water [44]. The sample geometry is free of extraneous high refractive index components, is compatible with arbitrarily small emitters, and provides non-optical ground truth axial positions of emitters. We show that our experimental geometry can be used to calibrate the depth-dependence of PSFs by using the open aperture (OA) PSF, astigmatic PSF [45], and double-helix (DH) PSF [46]. We report and experimentally investigate the nonlinear scaling of the nominal focal plane and axial emitter position within a few microns of the interface and compare the experimental results to theoretical calculations. Our experiments reveal that simple approximations of the focal shift that are traditionally used in the field do not suffice to describe the true behavior of the system at the length scale common to most single-molecule imaging. We discuss the axial biases resulting from various approximations and suggest further theoretical analysis in the future.

## 2. Experimental methods

### 2.1 Sample preparation protocol

Fluorescent beads (Molecular Probes, F8789, dark red fluorescent, 660/680) with a diameter of 40 nm were diluted to a final concentration of ~2-30 pM in water. The solution was spin-coated onto plasma-etched glass coverslips (Schott Nexterion #1.5H, Applied Microarrays) with a refractive index of $n_g = 1.525$. Next, the bottom surface of each coverslip was coated with an optical cleaning polymer (First Contact, Photonic Cleaning Technologies), which formed a removable layer (Fig. 1). The coverslips were then dip-coated in a solution of the polymer MY-133-MC (MY Polymers, Israel) diluted to 5% (v/v) in the fluorosolvent Novec 7100 (3M). The refractive index of this polymer was measured with a refractometer (Abbe-3L, Bausch & Lomb) to be $n_p = 1.330(3)$, closely matching the value of $n_w = 1.333(1)$ measured for nanopure water. Dipping was performed at 1.75 mm/s with a linear translation stage (MTS50-Z8, Thorlabs), with an incubation period of 30 s before withdrawal from the polymer solution. After the initial dip-coating, a second dip-coating was performed 5 min later, and a third 60 min later. In between dip-coatings, the samples were cured in air at room temperature (~21°C). After the third dip-coating, the optical cleaning polymer was removed from the bottom surface. This protocol produced a flat polymer layer approximately 125 nm thick on the top surface of the coverslip.

Next, up to 16 additional polymer layers were added to the coverslip via spin-coating of 50 µL of the 5% polymer solution (30 s at 10,000 rpm). Each layer was allowed to cure for 60 min, and each layer was found to add approximately 250 nm of total polymer thickness after curing (see Appendix A).

Prior to imaging, each sample was plasma-etched in Ar and 50 µL of poly-L-lysine (0.01% (w/v) in water, Sigma-Aldrich) was spin-coated onto the polymer surface. This was found to promote adhesion of beads to the polymer surface while having no measurable effect on polymer thickness. A small chamber (SA8R-1.0 SecureSeal Hybridization Chamber, Grace Bio-Labs) was adhered to the top of the polymer surface and filled with a ~4-15 pM solution of the same beads added to the coverslip surface. After a ~20 min incubation period, the chamber was washed ~10 times with 200 µL aliquots of water to remove unadhered beads.

### 2.2 Polymer thickness measurements

Prior to optical measurements, each sample was cut through the polymer 3-4 times with a scalpel (No. 11, Feather) so that the glass surface was exposed. The thickness of the polymer layer on each coverslip was then measured via stylus profilometry (Dektak XT, Bruker) across each cut at many locations throughout the coverslip. Each coverslip was profiled a total of ~30 times to assess uniformity of the polymer thickness in regions to be imaged (Appendix A), which was found to be ~0.5-2%.

To ensure that the polymer is locally flat and that its thickness is not changed by the addition of water, control experiments were performed. Atomic force microscopy (BioScope Resolve, Bruker) of the polymer surface was performed before and after the addition of water to the polymer surface, both near a scalpel incision and in an undisturbed region (Appendix A).

### 2.3 Optical setup

Imaging was performed on an inverted microscope (IX71, Olympus) with a precision XYZ stage (PInano XYZ Piezo Stage and High Precision XY Microscope Stage, Physik Instrumente). The sample was pumped by a 647 nm laser (OBIS, Coherent), which was reflected into the microscope by a dichroic mirror (Di01-R405/488/561/635, Semrock) and focused onto the back aperture of the objective by a Köhler lens (f = 350 mm) to achieve a peak wide-field illumination intensity of ~1-10 W/cm$^2$. This low intensity minimized bleaching effects during imaging.

Fluorescence was collected by a high-NA oil-immersion objective (UPLSAPO100XO, 100x, NA 1.4, Super Apochromat, Olympus). The tube lens focused the emission onto the intermediate image plane of the microscope, which was relayed onto the final image plane by a 4f optical processing system consisting of two achromat lenses (f = 90 mm). Fluorescence was spectrally filtered (ET700/75m and ZET635NF, Chroma) before imaging onto an electron-multiplying Si CCD camera (iXon3 897, Andor). A schematic of the optical setup is shown in Fig. 2.

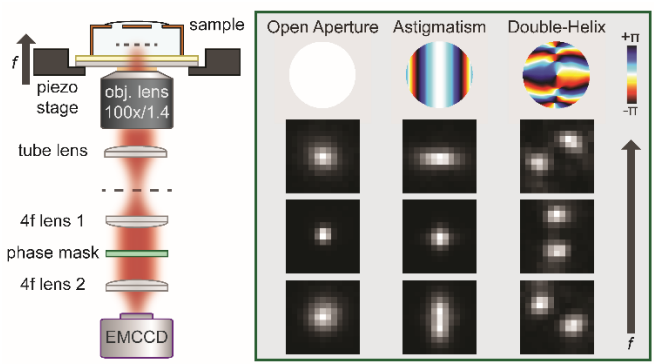

Fig. 2. Optical setup. Sample is mounted on a piezo stage and imaged with an oil-immersion objective lens (100x, NA 1.4). The image formed by the tube lens is relayed via a 4f system to the final image plane at the EMCCD camera. A phase mask is placed conjugate to the back focal plane. Inset shows phase patterns for the different PSFs used in radians (top) and examples of scaled images acquired at different focal positions (columns) from beads on the glass coverslip. Focal positions are offset by 500 nm (OA, astigmatism) or 750 nm (DH). Pixel size of the PSF images is 160 nm at the sample plane.

*2.4 Imaging procedure*

After mounting on the microscope, samples were placed in contact with immersion oil (Immersol 518 F, Carl Zeiss) with a measured refractive index of $n_o$ = 1.517(1). Samples were taped to the microscope stage and left for ~15-20 min prior to imaging in order to allow for equilibration and minimize stage and sample drift. The sample was scanned along the axial ($z$) direction in 50 nm steps using the piezoelectric stage. The scan was performed beginning with the focal plane below the glass-polymer interface and ending with the focal plane above the polymer-water interface, in order to observe both in-focus and out-of-focus images of the point spread functions in the two layers of beads in the sample. At each step in the scan, ten 50-ms frames were acquired continuously, and a wait time of 0.5 s was used after each movement of the stage to avoid effects of immersion oil and stage relaxation. During the wait time, the laser was blocked by a shutter to minimize total the total exposure time of the beads to the excitation beam. Importantly, conducting the scan by moving the microscope stage rather than the objective lens ensured that the objective lens would not move relative to the other optics and the location of the nominal focal plane would remain constant in the lab frame.

In order to average over slight heterogeneity in polymer thickness throughout the sample or drift during scans, a total of ~20-50 scans were acquired with a given PSF throughout each sample. Anomalous measurements could then be identified.

Engineered point spread functions were implemented by addition of phase-modulating optics in the emission path. The DH PSF was produced by placing a phase mask (686 nm design wavelength, Double Helix Optics) in the Fourier plane of the microscope. The astigmatic PSF was produced by placing a cylindrical lens (f = 1 m) ~70 mm before the final image plane.

## 3. Determination of focal shift

### 3.1 Localization analysis

Localization of the scanned PSF images was performed using common position estimators to determine the stage position at which each bead was "in focus." PSF images were localized using a symmetric Gaussian, asymmetric Gaussian, or symmetric double-Gaussian model for the OA, astigmatic, and DH PSFs, respectively (Appendix C). In all cases, model parameter values were determined using a least squares loss function.

For each PSF, the "in focus" image is defined differently. The OA PSF is in focus when its transverse width is minimized; the astigmatic PSF is in focus when its horizontal and vertical widths are equal; the DH PSF is in focus when its lobes have no horizontal offset (i.e., $\theta = 0°$). To determine these locations, calibration curves of PSF width (OA), PSF width ratio (astigmatic), and relative lobe angle (DH) are plotted against stage position and locally interpolated. For the astigmatic and DH PSF, a cubic spline interpolant was used, while the OA PSF calibration curve was locally fitted to a quadratic to find its minimum (see Fig. 4 for examples).

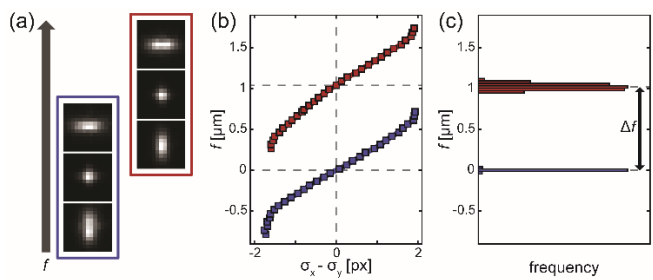

Fig. 3. Analysis protocol. (a) The stage is scanned to obtain images at different focal positions. Emitters at the glass-polymer interface (blue) and polymer-water interface (red) are each moved through their in-focus image (center). Images are spaced by 250 nm in $f$ and have a pixel size of 160 nm. (b) A calibration curve is measured from each emitter by localization. The focus criterion (here, difference between horizontal and vertical Gaussian widths on the abscissa) is plotted against nominal focal position $f$ to determine the $f$ position of the in-focus emitter. (c) Histograms of $f$ for many emitters at the glass surface (blue) and the polymer-water interface (red) are plotted. The change in nominal focus $\Delta f$ is determined from the difference between the means of the two histograms. Focal shift is calculated from $t/\Delta f$, where $t$ is the polymer thickness (here, 646 nm). Histogram bin size is 30 nm.

### 3.2 Focal shift calculation

From each scan, a calibration curve was generated for each bead in the field of view and used to determine the position of the piezo stage that yields an in-focus image of the bead (Fig. 3). Beads were located either at the glass-polymer (bottom) interface or the polymer-water (top) interface, so the aggregated results across many scans yielded two histograms for each polymer sample. Following rejection of bad fits (due to e.g. low signal-to-noise ratio or overlap with nearby emitters), the distance $\Delta f$ that the stage must be moved to translate the focus from the bottom to the top layer of emitters could be determined. To account for tilt of the coverslip, in each scan emitters on the bottom layer were fitted to a plane, and interpolation of the planar surface was used to determine the value of $\Delta f$ for each emitter in the top layer. The aggregated values of $\Delta f$ from all fields of view imaged within a sample were used to determine a single value of $\Delta f$.

The thickness of the polymer layer reports on the displacement of the actual focal plane during the scan. For a given polymer thickness $t$, the focal shift is then defined as the ratio $t/\Delta f$. This calculation was performed for each polymer sample and each of the three PSFs individually by aggregating many surface profilometry measurements and bead scans in many different fields of view in order to obtain a robust estimate of the focal shift from each sample.

## 4. Simulation of focal shift

### 4.1 Theoretical point spread function

In order to compare the results of our experiments to theoretical calculated determinations of the focal shift, the experimental geometry was simulated by generating the PSF shape following the established vectorial diffraction theory for image formation in high-NA microscopy [47,48]. The treatment begins by considering a monochromatic point dipole emitter with polar angle $\Theta_D$ and azimuthal angle $\Phi_D$, whose dipole orientation is described by the vector $\hat{\boldsymbol{\mu}} = [\sin\Theta_D \cos\Phi_D, \sin\Theta_D \sin\Phi_D, \cos\Theta_D]^T$ and whose emission wavelength is denoted by $\lambda$. The far field Green's tensor formalism [48-50] can be used to determine the electric field at the back focal plane (BFP with radial coordinates $\rho$, $\varphi$) of the microscope due to the toroidal emission pattern of the dipole from (see Appendix B for the explicit Green's tensor)

$$\boldsymbol{E}_{BFP}(\rho,\varphi \,|\, \hat{\boldsymbol{\mu}}, \lambda) \propto \boldsymbol{G}_{BFP}(\rho,\varphi \,|\, \lambda)\hat{\boldsymbol{\mu}}. \tag{1}$$

The above expression enables calculation of the electric field in the BFP due to a dipole located at the water-glass interface and in the focal plane of the microscope ($z = f = 0$). In our experiments, the nominal focal position may be located away from the coverslip by a distance $f \neq 0$ and the emitter may be located at a height of $z > 0$ away from the coverslip. Additionally, a phase mask may be added to produce engineered PSFs. These effects are incorporated into the model by means of a phase factor applied in the BFP [8], such that the final electric field has the form

$$\boldsymbol{E}_{BFP}(\rho,\varphi \,|\, \hat{\boldsymbol{\mu}}, \lambda) \propto \exp\left\{i\frac{2\pi}{\lambda}\left[z\sqrt{n_w^2 - (n_o\rho)^2} - f\sqrt{n_o^2 - (n_o\rho)^2} + \Psi_m(\rho,\varphi \,|\, \lambda)\right]\right\} \boldsymbol{G}_{BFP}(\rho,\varphi \,|\, \lambda)\hat{\boldsymbol{\mu}} \tag{2}$$

in which positive values of $z$ and $f$ represent displacements of the emitter and nominal focal plane into the imaging (aqueous) medium, respectively, and $\Psi_m$ represents the phase mask. The paraxial theory of Fourier optics then gives us the electric field in the image plane (denoted by primed coordinates) via a scaled Fourier transform according to

$$\boldsymbol{E}_{IP}(\rho',\varphi' \,|\, \hat{\boldsymbol{\mu}}, \lambda) \propto \int_0^{2\pi}\int_0^{\rho_{max}} \boldsymbol{E}_{BFP}(\rho,\varphi \,|\, \hat{\boldsymbol{\mu}}, \lambda) \exp\left[i\frac{2\pi}{\lambda f_{TL}}\rho\rho'\cos(\varphi'-\varphi)\right]\rho\,d\rho\,d\varphi \tag{3}$$

where $f_{TL}$ is the tube lens focal length [51]. The intensity in the image plane for a dipole PSF is then given by $I_{IP}(\rho',\varphi' \,|\, \hat{\boldsymbol{\mu}}, \lambda) = |\boldsymbol{E}_{IP}(\rho',\varphi' \,|\, \hat{\boldsymbol{\mu}}, \lambda)|^2$.

A fluorescent bead is modeled as a ball of rigid dipole emitters sampling random orientations on the unit sphere [52]. For a bead of infinitesimally small size, i.e. a true point source, this entails weighting the contribution of each dipole by its absorption probability. For a given excitation electric field $\boldsymbol{E}_{exc}$, the absorption probability for a dipole is $|\boldsymbol{E}_{exc}\cdot\hat{\boldsymbol{\mu}}|^2$ such that the final bead PSF is given by

$$I_{IP}^{bead}(\rho',\varphi' \,|\, \lambda) = \iint d\hat{\boldsymbol{\mu}} \,|\boldsymbol{E}_{exc}\cdot\hat{\boldsymbol{\mu}}|^2 I_{IP}(\rho',\varphi' \,|\, \hat{\boldsymbol{\mu}}, \lambda). \tag{4}$$

To incorporate the effect of finite emitter size, one approach is to convolve the result of Eq. (4) with a sphere of radius $r_{bead}$ [53], noting that the bead contains a distribution of $z$ positions and thus requires a depth-variant convolution. Additionally, the spectrum of emitted fluorescence from the bead can be incorporated into the model by integrating the contributions from the constituent wavelengths incoherently. The effect of these changes to the model on the simulated focal shift curves is subtle and can be reasonably well-approximated by the less computationally intensive approach of blurring $I_{IP}^{bead}(\rho',\varphi'|\lambda)$ with a Gaussian kernel [54,55]. The effects of this approximation on other aspects of PSF simulation or localization are not studied quantitatively here. We further note that larger beads (e.g. 100 nm or greater) are more susceptible to effects due to blurring. The 40 nm beads used in this study have a minimal contribution from blurring while remaining bright and stable during imaging.

### 4.2 Phase-retrieved point spread function

In order to refine the PSF model from the previous section to improve the correspondence between the experimental and simulated PSFs, a phase retrieval step was incorporated in the PSF modeling. This was accomplished with an algorithm demonstrated previously [53,54], which has been shown to provide good estimates of phase aberrations present in experimental PSFs and to improve 3D localization precision. Briefly, the algorithm uses a stack of PSF images of an emitter at the coverslip ($z = 0$) and at different $f$ positions, similar to the ones acquired throughout this paper, to estimate an additional phase correction term to multiply the electric field in the back focal plane. This is accomplished using maximum likelihood estimation with a Poisson noise model. The unknown phase term is decomposed into a sum of Zernike polynomials $Z_j(\rho,\varphi)$, and the modified electric field in the back focal plane becomes

$$\boldsymbol{E}_{BFP}^{PR}(\rho,\varphi|\hat{\boldsymbol{\mu}},\lambda) = \exp\left[i\sum_j c_j Z_j(\rho,\varphi)\right] \boldsymbol{E}_{BFP}(\rho,\varphi|\hat{\boldsymbol{\mu}},\lambda). \quad (5)$$

The function $\boldsymbol{E}_{BFP}^{PR}(\rho,\varphi|\hat{\boldsymbol{\mu}},\lambda)$ is then used in place of $\boldsymbol{E}_{BFP}(\rho,\varphi|\hat{\boldsymbol{\mu}},\lambda)$ in Eq. (3) and the downstream calculations of the bead PSF. The task of the estimation procedure is to determine the values of the coefficient vector $\boldsymbol{c} \triangleq \{c_j\}$ that maximize the likelihood of observing the measured PSF data. This is done by minimizing the negative logarithm of the likelihood function for Poisson noise in the detected photon counts, i.e.

$$-\log \mathcal{L}\left(\boldsymbol{c};\{\tilde{X}_{k,s}\}\right) = \sum_k \sum_s (X_{k,s}|\boldsymbol{c}) - \tilde{X}_{k,s} \log(X_{k,s}|\boldsymbol{c}) + \log(\tilde{X}_{k,s}!) \quad (6)$$

in which $\tilde{X}_{k,s}$ represents the measured photon count in pixel $k$ of image $s$ of the experimental PSF stack, and $(\tilde{X}_{k,s}|\boldsymbol{c})$ is the simulated photon count in pixel $k$ of image $s$ of the simulated PSF stack with aberration vector $\boldsymbol{c}$.

### 4.3 Focal shift simulations

In simulating focal shift measurements, we strove to produce a realistic model of the experiment. The theoretical focal shift determination for a sample with polymer thickness $t$ begins with a bead simulated at the water-glass interface by approximating it as an isotropic, monochromatic point source located at a height of $z = r_{bead}$ (20 nm, the bead radius) above the coverslip. Similarly, a second bead is approximated the same way but located at a height of $z = t + r_{bead}$. These two beads are simulated at various positions of the nominal focus, and each simulated image is localized in the same way as the experimental data, using the appropriate PSF model function (Appendix C). The parameters of the fits yield calibration curves for the two beads, from which the nominal focus value at which each bead is in focus is determined.

The difference between the two nominal focus position reports on the Δ$f$ in the simulation, and division of $t$ by this number yields the corresponding focal shift.

## 5. Results and discussion

### 5.1 PSF calibration curves

Each scan of the nominal focus through a sample provides images of emitters at two different axial positions: at the top and bottom surfaces of the polymer. After sorting the emitter images into the two layers, the PSF can be localized in each layer as a function of the nominal focal position. This can be displayed as calibration curves like those in Fig. 4 below. In each image, the emitter is localized as described in Sec. 3.1 and its PSF width (OA), difference between horizontal and vertical PSF widths (astigmatism), or lobe angle (DH) is extracted and plotted on the ordinate.

The results (Fig. 4) quantitatively describe the effects of refractive index mismatch on PSFs of emitters at various depths. It can be seen that each of the three PSFs loses contrast with increasing depth, as the slope of the calibration curve decreases. Additionally, a depth-dependent asymmetry can be seen with respect to the "in-focus" position, corresponding to $\delta f = 0$. When the focal plane is displaced away from the in-focus position and toward the coverslip ($\delta f < 0$), with increasing $z$ the PSFs lose contrast more quickly and the calibration curves change more strongly with respect to the in-focus emitter position.

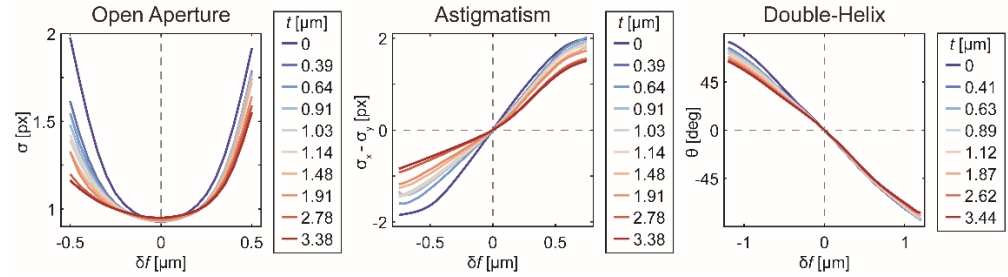

Fig. 4. Depth-dependent calibration curves. In each panel, calibration curves measured in samples of different polymer thickness are plotted. The abscissa indicates displacement of the piezo stage from the focal position, i.e. each curve is shifted so its focus occurs at $\delta f = 0$. Criteria for determining focal plane position are minimum PSF standard deviation (left, OA), difference between horizontal and vertical PSF standard deviations (middle, astigmatism), and PSF lobe angle (right, DH).

### 5.2 Focal shift

Whereas in Sec. 5.1 the depth-varying shape of the different PSFs was explored at different amounts of defocus using the calibration curves, we now turn our attention to the shift of the "in-focus" position ($\delta f = 0$ in Fig. 4) with axial emitter depth $z$. The distance between the observed in-focus positions of beads at the glass-polymer and polymer-water interfaces is given by Δ$f$. The relationship between Δ$f$ and $z$ is often approximated to be linear, i.e. $z = c\Delta f$, implying a constant focal shift. Importantly, both simulations and experiments demonstrate that the focal shift in fact varies with emitter depth (Fig. 5).

In studying the theoretical predictions (solid curves), we note that the three PSFs yield relatively similar experimental results, but that in all cases agreement between experiment and theory is limited to ~2 or more microns away from the coverslip. As the polymer thickness approaches 0, the deviation between experiment and simulation becomes greater, with focal shift being overestimated by close to ~15% at the lowest data point of ~390 nm. Importantly, the region closest to the coverslip is precisely where most, but not all, single-molecule imaging is performed. Evidently, it is the region in which our prediction of the true focal shift is poorest.

We note that the simulations in Fig. 5 use phase-retrieved PSFs created as outlined in Sec. 4.2 rather than the purely theoretical ones described in Sec. 4.1. This is because phase retrieval significantly improves the correspondence between theory and experiment, especially for engineered PSFs [30,31,41,53-58]. A comparison between the simulated focal shifts before and after the addition of phase-retrieved aberrations is shown in Appendix D.

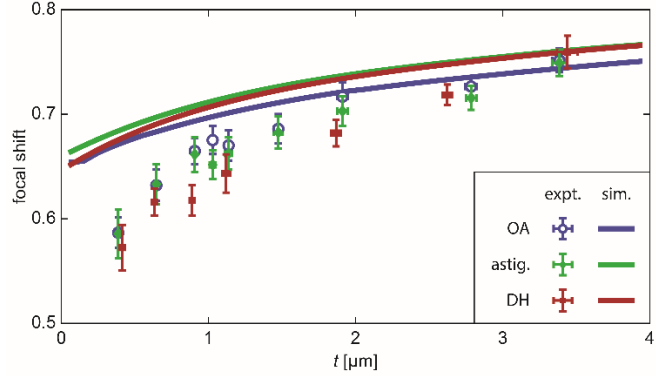

Fig. 5. Depth-dependence of focal shift for emitter at $z$ above the coverslip. Data points are measured using samples of polymer thickness corresponding to the axial position. Error bars indicate standard deviation (Appendix E). Solid curves show simulated focal shift based on a phase-retrieved PSF. Colors indicate OA (blue), astigmatic (green), or DH (red) PSFs.

### *5.3 Axial localization bias*

To demonstrate the effect of poorly-corrected focal shift on experimental data, a simulation was performed using a fourth-order polynomial fit to the experimental focal shift data (functional form given in Appendix F) as the ground truth. The polynomial fit gives a "true" (empirical) value of the focal shift at each emitter axial position from 0 to 4 microns above the coverslip. We will use this fit to deduce an effective bias in axial position resulting from incorrect modeling of the focal shift in the following way. Starting with a "true" $z$ position, division of the emitter axial position by the "true" focal shift yields $\Delta f$ at each axial position. Next, to convert $\Delta f$ to an estimated $z$ position, one must choose one of the theoretical focal shift curves or the constant focal shift model. Of course, choosing the polynomial curve that generated the ground truth yields no bias in this simulation. However, for other choices of focal shift curves, the estimated z position may be biased relative to the true $z$ position. In Fig. 6 below, the axial bias, $z_{est} - z_{true}$, is plotted in the top row, and its percentage of $z_{true}$ is plotted in the bottom row.

Four different corrections are considered: the commonly used constant focal shift of 0.8 (dark blue) or 0.65 (light blue) and depth-dependent focal shift based on unaberrated (yellow) or phase-retrieved (red) PSF simulations. While the focal shift estimates from the phase-retrieved model results in the lowest axial bias overall, depth-varying axial bias is observed for all three PSFs. The percentage error becomes especially significant within about a micron of the coverslip, precisely in the region where the experimental focal shift curves deviate most from experiment. In particular, for the OA PSF, the phase-retrieved correction achieves quite low percentage error for $z$ above ~1.5 μm, yet no model achieves low bias for the other two PSFs. Our analysis implies that even with the best models available, the location of the actual focal plane is incorrectly estimated, especially near the coverslip, leading to axially-distorted localization data.

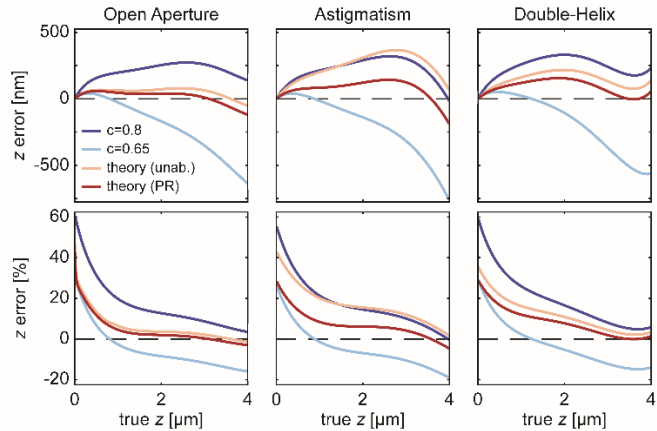

Fig. 6. Simulations of axial bias. Each plot shows absolute *z* error (top row) or percentage *z* error relative to true position above the coverslip (bottom row). Colors indicate different error-correction approaches. Simulations are based on the experimental focal shift data for each PSF fitted with a 4$^{th}$ order polynomial. Dashed lines at 0 bias are included to guide the eye.

## 6. Conclusions

In this paper, we have demonstrated an experimentally useful model for water-glass refractive index mismatch in 3D single-emitter localization studies. The experimental sample, based on layers of polymer matched to the refractive index of water, lends itself to careful study of the depth-dependence of PSFs near the water-glass interface. We used an array of samples with different polymer thicknesses to calculate depth-dependent calibration curves for the OA, astigmatic, and DH PSFs, quantitatively showing the asymmetry and loss of contrast introduced by the refraction at the glass surface. Such calibration curves containing ground truth position measurements will be a useful tool both for performing accurate 3D localization microscopy as well as a tool for studying the effects of refractive index mismatch on the PSF in various optical systems more generally. Additionally, this method provides a benchmark for alternative approaches.

We also demonstrated that a complicated, depth-varying relationship exists between the actual and nominal focus positions in our index-mismatched high-NA imaging system. We showed experimental measurements of the depth-dependent focal shift near the water-glass interface for the first time, demonstrating qualitative agreement with our theoretical calculations as well as previous theoretical work [38].

We also showed, however, that the accuracy of theoretical predictions of the focal shift is limited. Phase retrieval was able to improve agreement with experiment, especially far from the coverslip, but within the first micron from the coverslip (where many single-molecule localization experiments take place), our measurements deviate from the simulations, suggesting that translation of the actual focal plane behaves in a complicated way near the water-glass interface.

With a simulation, we calculated the axial localization bias incurred by several approaches to correcting the focal shift. It is clear that the descriptions of focal shift by vectorial diffraction theory, which reveal its depth-dependent nature, are superior to the linear scaling approximation (i.e. constant focal shift) used frequently by researchers. Calibration of this phenomenon may be the more desirable approach until a more complete simulation of the PSF can be produced which successfully describes the full focal shift curve in agreement with experiments. It is clear that phase retrieval will play a role in correct modeling; whether and how additional modifications to the theoretical PSF model should be made remains to be seen. Effects to be considered include the possibility of a non-ideal transfer function of the microscope objective and subtle polarization effects from supercritical light.

Importantly, an accurate PSF model contains features like focal shift implicitly within it. Although calibration of the depth-dependent PSF shape through simple Gaussian models as performed here is an attractive and popular option for expediting 3D localization microscopy data analysis, we believe that accurate simulation of the PSF will empower researchers to be rid of such abstractions. We hope that the experimental model developed in this work can provide a useful platform for further studying and improving upon existing PSF models. An accurate description of this experimental geometry lends itself not only to 3D localization fluorescence microscopy but also to other imaging modalities in which a water-glass refractive index mismatch geometry is present such as light sheet, interferometric scattering, deconvolution, and confocal microscopies.

**Appendix A**

Each sample prepared in this study had a single polymer thickness, allowing for focal shift measurement at one depth above the glass coverslip (height above coverslip in an inverted microscope). After the polymer was prepared and fully cured atop the glass coverslip, stylus profilometry was performed to check for sample quality and measure the polymer thickness. Each sample was cut 3-4 times with a scalpel and each cut was profiled in ~8-12 different places by performing linear profilometry scans perpendicular to the cut axis (Fig. 7). The cut was positioned roughly in the center of each 0.5 mm scan. Regions on either side of the cut were fit to a line to account for sample tilt and set to an axial position of zero. Since the cut extended all the way to the glass, the polymer thickness at the location of the scan then corresponded to the negative of the minimum axial position at the cut location. The mean cut depth across all scans within a sample was assigned as the polymer thickness. The standard deviation of cut depth measurements is reported in the x-axis error bars in Fig. 5. Importantly, outliers due to e.g. sample damage could be detected and those regions of the sample were avoided during imaging.

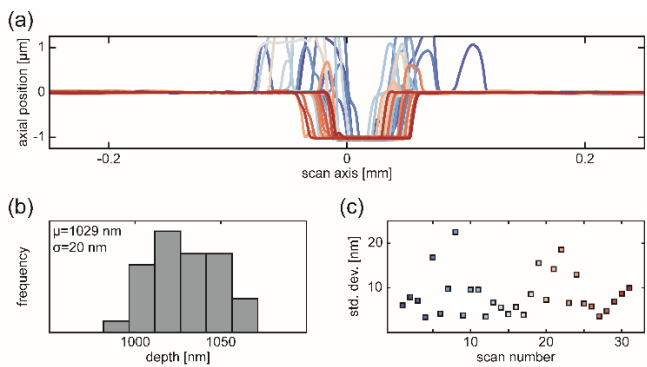

Fig. 7. Surface profilometry. (a) Line scans 0.5 mm in length were made across incisions at different locations in the sample. Very close to the cut, the polymer surface was distorted upward, but undisturbed regions on the top of the polymer and at the glass surface were flat. (b) Histogram of depths determined by maximum minus minimum measured axial position after a linear fit to the flat parts of each scan. The mean reports the thickness used as a measure of emitter axial position during imaging. Standard deviation is the horizontal error bar in Fig. 5. (c) Standard deviation in axial position along the undisturbed portion of the polymer in each scan shows flatness.

Flatness of the samples was checked by measuring the standard deviation of axial positions within 0.1 mm of the edge of each scan. We note that this provides an upper bound on sample flatness, since it does not account for large-scale warping of the coverslip.

In order to achieve different thicknesses in the 0-4 μm range, polymer solution was spin-coated onto existing samples as described in Sec. 2.1. Fig. 8 shows the polymer thickness as a function of number of polymer layers added. A linear fit reveals ~250 nm of polymer thickness is added with each spin-coated layer.

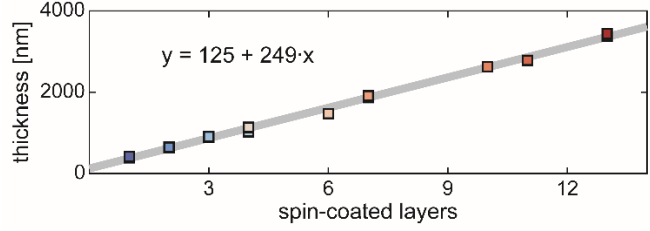

Fig. 8. Polymer layer thickness as a function of number of spin-coated layers. All samples included in Fig. 5 are plotted. A linear fit reveals that spin coating increased the polymer thickness by 249 nm per coat.

In order to be able to perform polymer thickness measurements in air prior to addition of water and fluorescence imaging, we used atomic force microscopy (AFM) to measure the polymer thickness before and after addition of water.

AFM was performed first in air near the edge of a scalpel cut (made as described in above). Water was added to the polymer surface, and after ~2 hours a second AFM scan was conducted in the same region of the sample but in the aqueous environment. The height of the polymer above the interface was determined from a histogram of the axial positions in each image, and essentially the same thickness $t$ was recovered in both cases (Fig. 9).

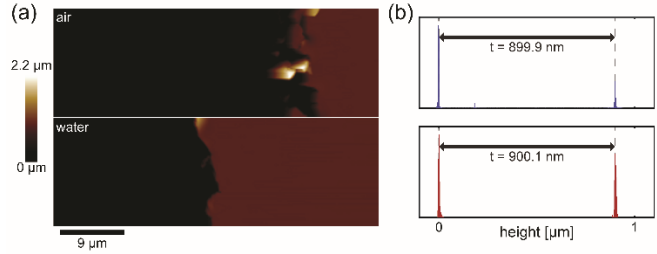

Fig. 9. Atomic force microscopy. (a) Micrographs in the same region of a polymer sample (dark red area), near a cut made with a scalpel (black area). Following the scan in air (top), the sample was immersed in water for ~2 hrs before acquisition of a scan in water (bottom). (b) Histograms of measured heights show peak separations are unchanged by the addition of water to the top of the sample.

## Appendix B

The Green's tensor used for vectorial image calculations is given by

$$G_{BFP}(\rho,\varphi|\lambda) = \frac{n_o}{n_w}\sqrt{1-\rho^2}\begin{bmatrix} \alpha\sin^2\varphi + t^p(\rho)\cos^2\varphi & \beta & \gamma\cos\varphi \\ \beta & \alpha\cos^2\varphi + t^p(\rho)\sin^2\varphi & \gamma\sin\varphi \\ 0 & 0 & 0 \end{bmatrix} \quad (7)$$

where $\alpha \triangleq t^s(\rho)/\sqrt{1-(n_o\rho/n_w)^2}$ , $\beta \triangleq \sin\varphi\cos\varphi\left[t^p(\rho)-t^s(\rho)/\sqrt{1-(n_o\rho/n_w)^2}\right]$ ,

$\gamma \triangleq -n_o t^p(\rho)\rho\cos\varphi/n_2\sqrt{1-(n_o\rho/n_w)^2}$ , and $t^{s,p}(\rho)$ are the Fresnel transmission coefficients for s- and p-polarized light with an incidence angle of $\theta_w = \sin^{-1}(n_o\rho/n_w)$.

## Appendix C

Position estimators based on Gaussian functions were used for all three PSFs used in this work.

For the open aperture PSF, the model function used was a symmetric Gaussian with free parameters $\{\theta_x, \theta_y, \theta_\sigma, \theta_A, \theta_b\}$ and the functional form

$$M(x', y') = \theta_b + \theta_A \exp\left\{-\frac{\left[(x'-\theta_x)^2 + (y'-\theta_y)^2\right]}{2\theta_\sigma^2}\right\}. \tag{8}$$

For the astigmatic PSF, an asymmetric Gaussian was used, with free parameters $\{\theta_x, \theta_y, \theta_{\sigma x}, \theta_{\sigma y}, \theta_A, \theta_b\}$ and the functional form

$$M(x', y') = \theta_b + \theta_A \exp\left[-\frac{(x'-\theta_x)^2}{2\theta_{\sigma x}^2} - \frac{(y'-\theta_y)^2}{2\theta_{\sigma y}^2}\right]. \tag{9}$$

For the DH PSF, a symmetric double-Gaussian was used, with free parameters $\{\theta_{x1}, \theta_{y1}, \theta_{x2}, \theta_{y2}, \theta_{\sigma 1}, \theta_{\sigma 2}, \theta_{A1}, \theta_{A2}, \theta_b\}$ and the functional form

$$M(x', y') = \theta_b + \sum_{i=1,2} \theta_{Ai} \exp\left\{-\frac{\left[(x'-\theta_{xi})^2 + (y'-\theta_{yi})^2\right]}{2\theta_{\sigma i}^2}\right\}. \tag{10}$$

In each case, localization was performed by a least squares fit of the model to the corresponding data.

**Appendix D**

The theoretical model of the PSF described in Sec. 4.1 assumes a perfect imaging system. In practice, limitations in the manufacturing and alignment of optical components within high-NA microscopes limits the accuracy of the theoretical model. In order to improve the correspondence between the experimental and simulated PSFs, we use phase retrieval to estimate system aberrations as described in Sec. 4.2. It has been shown that phase-retrieved PSFs have an improved capacity as models of the PSF to be used for localization microscopy (in place of e.g. the Gaussian approximations above, or unaberrated PSF models produced by "pure" theory). In addition, we show in Fig. 10 that phase-retrieved PSF models produce improved simulations of the focal shift relative to the predictions of the unaberrated diffraction theory.

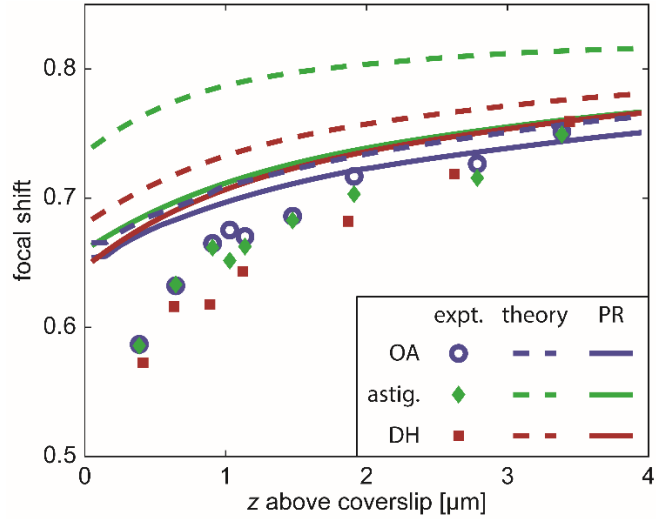

Fig. 10. Role of phase retrieval. Experimental data is replotted from Fig. 4 (error bars omitted for clarity). Solid lines show simulated focal shift based on phase-retrieved PSF models (replotted form Fig. 4). Dashed lines show focal shift simulations based on unaberrated PSF models. For each PSF, addition of phase aberrations to the PSF model improved correspondence between simulations and experimental data, particularly for thick samples and for engineered PSFs.

## Appendix E

In order to estimate the error in emitter axial position, multiple measurements of the polymer thickness were made at different locations throughout each sample and the standard deviation was determined, yielding a thickness estimate $t \pm \sigma_t$ for each sample. Likewise, displacement of the nominal focal plane between the two layers of beads in each polymer sample was tracked using a piezo stage, and the standard deviation of those distance estimates yielded, for each coverslip, $\Delta f \pm \sigma_{\Delta f}$. The focal shift, calculated from the ratio $t/\Delta f$, has error bars given by the standard deviation $\sigma_{t/\Delta f} = (t/\Delta f)\sqrt{(\sigma_t/t)^2 + (\sigma_{\Delta f}/\Delta f)^2}$.

## Appendix F

A fourth-order polynomial fit to the data in Fig. 4 was used in Sec. 5.3 to perform a simulation of axial biases incurred by several approximations of the focal shift. This polynomial has the form

$$p_4(z) = -0.00534z^4 + 0.04652z^3 - 0.14408z^2 + 0.23786z + 0.50021. \qquad (11)$$

## Funding

U. S. National Institute of General Medical Sciences (NIGMS) (R35GM118067); Xu Family Foundation Stanford Bio-X Interdisciplinary Graduate Fellowship; National Science Foundation (NSF) (ECCS-1542152).

## Acknowledgements

We thank Peter Dahlberg, Katherine Liu, and Anna-Karin Gustavsson for helpful discussions and Marcin Walkiewicz for guidance regarding atomic force microscopy measurements. Part of this work was performed at the Stanford Nano Shared Facilities (SNSF), supported by the National Science Foundation.